\documentclass[aps,prd,twocolumn,showpacs,floatfix,preprintnumbers,amsfont,amsmath,amssymb,nofootinbib, superscriptaddress]{revtex4-1}

\usepackage{color}
\usepackage{hyperref}
\usepackage[normalem]{ulem}
\usepackage{lipsum}
\usepackage{url}
\usepackage{float}
\usepackage{ctable}
\usepackage{graphicx}
\usepackage[font=small, justification=raggedright, format=plain
  ]{caption} 
\usepackage{subcaption}
\usepackage{mathtools}
\usepackage{bm}
\usepackage{siunitx}
\usepackage[utf8]{inputenc}

\newcounter{magicrownumbers}

\newcommand{\sk}[1]{}

\newcommand{\reffig}[1]{Fig.~\ref{fig:#1}}

\newcommand{\tabline}{\specialrule{.1em}{.05em}{.05em}}

\newcommand{\rmd}{{\rm d}}
\newcommand{\pastro}{\ensuremath{p_{\rm astro}}}

\newcommand{\tM}{{\tilde M}}
\newcommand{\tz}{{\tilde z}}

\def\chieff{\chi_{\rm eff}}

\newcommand{\be}{\begin{equation}}
\newcommand{\ee}{\end{equation}}
\newcommand{\ba}{\begin{eqnarray}}
\newcommand{\ea}{\end{eqnarray}}

\newcommand{\apjl}{Astrophys. J. Lett.}
\newcommand{\aap}{Astron. Astrophys.}

\newcommand{\mnras}{Mon. Not. R. Astron. Soc.}

\newcommand{\jcap}{Journal of Cosmology and Astroparticle Physics}

\allowdisplaybreaks

\begin{document}

\title{Search for Lensed Gravitational Waves Including Morse Phase Information: \\ An Intriguing Candidate in O2}
\author{Liang Dai}
\email{ldai@ias.edu}

\affiliation{\mbox{School of Natural Sciences, Institute for Advanced Study, 1 Einstein Drive, Princeton, NJ 08540, USA}}
\affiliation{\mbox{Department of Physics, University of California, Berkeley, 366 LeConte Hall, Berkeley, CA 94720, USA}}

\author{Barak Zackay}
\affiliation{\mbox{School of Natural Sciences, Institute for Advanced Study, 1 Einstein Drive, Princeton, NJ 08540, USA}}
\affiliation{\mbox{Dept. of Particle Physics \& Astrophysics, Weizmann Institute of Science, Rehovot 76100, Israel}}
\author{Tejaswi Venumadhav}
\affiliation{\mbox{School of Natural Sciences, Institute for Advanced Study, 1 Einstein Drive, Princeton, NJ 08540, USA}}
\affiliation{\mbox{Department of Physics, University of California at Santa Barbara, Santa Barbara, CA 93106, USA}}
\affiliation{\mbox{International Centre for Theoretical Sciences, Tata Institute of Fundamental Research, Bangalore 560089, India}}
\author{Javier Roulet}
\affiliation{\mbox{Department of Physics, Princeton University, Princeton, NJ, 08540, USA}}

\author{Matias Zaldarriaga}
\affiliation{\mbox{School of Natural Sciences, Institute for Advanced Study, 1 Einstein Drive, Princeton, NJ 08540, USA}}

\date{\today}


\begin{abstract}

We search for strongly lensed and multiply imaged gravitational wave signals in the second observing run of Advanced LIGO and Advanced Virgo (O2). 
We exploit a new source of information, the so-called Morse phase, which further mitigates the search background and constrains viable lenses. 
The best candidate we find is consistent with a strongly lensed signal from a massive binary black hole (BBH) merger, with three detected images consisting of the previously catalogued events GW170104 and GW170814, and a subthreshold trigger, GWC170620. 
Given the number of BBH events detected so far, we estimate an overall false alarm probability $\sim 10^{-4}$ for the observed high degree of parameter coincidence between the three events.
On the flip side, we measure the Morse phase differences which suggest a complex and atypical lens system, with at least five images including a magnified image at a local maximum of the Fermat potential. 
The low prior probability for multiple lensed images and the amount of fine tuning required in the lens model reduce the credibility of the lensing hypothesis.
The long time delays between lensed images point toward a galaxy cluster lens with an internal velocity dispersion $\sigma \sim \SI{650}{\kilo\meter/\second}$, and the observed strain amplitudes imply a likely range $0.4 < z \lesssim 0.7$ for the source redshift.
We provide an error ellipse of $\sim 16\,{\rm deg}^2$ for the sky location of the source together with additional specific constraints on the lens-host system, and encourage follow-up efforts to confirm or rule out any viable lens.
If this is indeed a lensed event, successfully pinpointing the system would offer a unique opportunity to identify the host galaxy of a BBH merger, and even localize the source within it.

\end{abstract}

\maketitle

\section{Introduction}

In general relativity, gravitational waves (GWs) are subject to gravitational lensing by intervening masses along the wave trajectory from the source to the observer (in an entirely analogous manner to electromagnetic waves~\cite{1996PhRvL..77.2875W}). In the strong lensing regime, this results in the formation of multiple images of a single source. It was suggested earlier~\cite{2013JCAP...10..022P, Biesiada:2014kwa} and has become clear since the discovery of the first binary merger, GW150914, by the LIGO and Virgo collaboration~\cite{GW150914} that the most promising case of lensing of GWs would involve signals from merging binary black holes (BBHs) magnified by intervening galaxy or galaxy cluster lenses~\cite{DaiGWLensingPopulation, NgPreciseLIGOLensingPredictions, Li:2018prc, OguriGWLensing, RobertsonLensingPredictions}, because these sources are routinely detected from cosmological distances $z\gtrsim 0.3$. 

Identifying multiple lensed images of a BBH source will enable a search for the underlying lens--host system in the optical catalog of galaxy/cluster lenses based on information extracted from the GW observations. This opens up the possibility of pinpointing the host galaxy of a BBH merger, and even locating the source environment within the host galaxy~\cite{Hannuksela:2020xor}. Besides, the strong lensing by a macroscopic lens can enhance the chance that one or several images are further perturbed by substructures within the lens, producing wave diffraction imprints that may shed light on the elusive sub-galactic dark matter halos devoid of luminous matter~\cite{Dai:2018enj} or stellar mass objects~\cite{HannukselaGWLensing, Diego:2019lcd}. Detection and follow-up study of lensed BBHs should foster the development of methodologies applicable to rarer lensed GW events with electromagnetic counterparts at future ground-based detectors. Precise characterization of the associated host-lens systems will enable new cosmological tests~\citep{liao2017precision, wei2017strongly, li2019constraining}. 

Motivated by the high values of the redshifted masses of the black holes in LIGO--Virgo BBH mergers, there have been suggestions that several BBH signals detected during the LIGO--Virgo O1 and O2 observing runs may have been multiple images of lensed sources~\cite{2018arXiv180205273B, 2018MNRAS.475.3823S, 2019arXiv190103190B}. On the other hand, theoretical calculations suggest that the observed rate of detected strongly lensed GW events during the O2 run should be small ($\sim 10^{-3}$--$10^{-2}$ per year), assuming standard populations of BBH sources and galaxy lenses~\cite{DaiGWLensingPopulation, NgPreciseLIGOLensingPredictions, Li:2018prc, OguriGWLensing, RobertsonLensingPredictions, ContigianiGWLensingEfficiency}. Unexpectedly high merger rates beyond the current detection horizon, at $z\sim 1$--$2$, can greatly enhance the apparent fraction of lensed events, but such scenarios will be subject to constraints on the stochastic background from the population of unresolved mergers~\cite{Mukherjee2020SGWBLensing, Buscichio2020SGWBLensing}. Previous analyses did not uncover significant evidence for either magnification or multiple images due to galaxy lensing among O1 and O2 BBH events~\cite{HannukselaGWLensing}, and certainly not all BBH events hitherto detected could have been strongly lensed~\cite{ConnorOGCLensingSearch}.

In the geometrical optics regime, we would interpret an isolated lensed image of a BBH signal as an unlensed event with different intrinsic masses and from a different redshift~\cite{DaiGWLensingPopulation, OguriGWLensing, ContigianiGWLensingEfficiency}. We can identify multiple lensed images in a set of distinct signals by looking for significant coincidences in the intrinsic source parameters and consistency between the sky localizations. 
The signals in the different images should also be consistent with a single source inclination and polarization angle. 
There is an additional subtlety with regards to the orbital phase: a geometrically lensed waveform acquires a phase shift, the so-called Morse phase, that equals $\pi/2$ times the number of negative eigenvalues of the lensing Jacobian matrix. 
This distinguishes images corresponding to minima, saddle points, and maxima of the Fermat potential~\cite{1986ApJ...310..568B}. Consequently, the inferred orbital phases of multiple images of a merger will differ by integer multiples of $\pi/4$~\cite{Dai:2016igl}. For single events, absolute orbital phases are poorly determined since they are degenerate with other extrinsic parameters. However, when considering the possibility that several events are lensed images of the same source, we can jointly infer the parameters for all signals by fixing the inclination and polarization angle to common values (that we marginalize over), but allowing independent orbital phases for different events. 
This procedure typically provides a good measurement of the difference in the orbital phases (though each of them is ill-determined due to the degeneracy mentioned above). These differences should have random values for unrelated events, but should be integer multiples of $\pi/4$ for multiple images of the same source. 
Previous searches for strongly lensed GW events did not include Morse phases in their criteria, and did not measure their values for their candidates~\cite{HannukselaGWLensing, AlvinLensingSearch, HarisGWLensingTechnique, ConnorOGCLensingSearch}. 
In this work, we demonstrate that the Morse phases provide crucial information that reduces the search background and informs us about any potential lens model.

Given the small prior probability for strong lensing, we require dramatic evidence to robustly associate a set of candidate multiple images with each other. 
For heavy (GW150914-like) BBH events observed by the two LIGO detectors with the typical signal-to-noise ratios (SNRs) achieved during the O2 run, it is challenging to establish that different signals are lensed images based on their consistency in intrinsic parameters and sky localization, unless the arrival time difference is shorter than a few hours~\cite{HarisGWLensingTechnique}. For lower mass systems, the detector-frame chirp masses are well measured, and hence give discriminating power to test the lensing hypothesis. 
However, these events come on average from shorter distances, and consequently the expected rate of lensed events (and hence the prior probability for lensing) is lower, due to which the lensing association is just as challenging to make~\cite{HannukselaGWLensing}. 
This makes past searches for lensed events virtually insensitive to genuine lensed events, as the standard Bayesian procedure for interpreting agreements in physical parameters would always prefer the unlensed hypothesis (with the exception of event pairs with very short time delays). In other words, at current detector sensitivities, it is unlikely to achieve a statistically significant association of pairs of events.

This limitation will be drastically reduced if we consistently detect events with three or more detectors in forthcoming observing runs, in which case the improved sky location and polarization information give us more discriminating power\cite{Pankow:2018phc, 2018LRR....21....3A}. 
Furthermore, next-generation instrumentation can improve the sensitivity at lower frequencies~\cite{Adhikari:2020gft} and increase the precision on all measured parameters of the signals, and enable the detection of low mass (say, chirp mass $\lesssim 10\,M_\odot$) BBH events from cosmological distances.
In this work, we do our best to exploit all available information given the sensitivity of GW data from O2 by searching for lensing association of more than two BBH events.

Surprisingly, our search yielded a very improbable candidate, with three BBH events associated as lensed images of a single source.
Two events, GW170814 and GW170104, have previously been detected in searches conducted by the LIGO/Virgo collaboration as well as by other independent groups~\cite{GWOSC, GW170104, GW170814, NitzCatalog, GWTC-1, O2CatalogIAS}. 
Previously in Ref.~\cite{HannukselaGWLensing}, this pair was disfavored due to its long time delay. 
For the same reason, this pair was not considered as a candidate in Ref.~\cite{ConnorOGCLensingSearch}, even though it had the best degree of parameter coincidence.
We show that two factors make this candidate interesting enough to warrant further investigation: Firstly, the Morse phase difference is consistent with one of the allowed values in geometrical lensing. Secondly, for low source redshifts, there is a significant relative contribution to the strong lensing optical depth from galaxy clusters, which naturally produce long time delays.
Given these considerations, we perform a targeted search for sub-threshold lensed images, restricting intrinsic parameters, sky location, and Morse phase shift to be compatible with association with GW170814 and GW170104. 
This targeted search uncovered a third BBH candidate, GWC170620, with high significance. 
This candidate did not pass the reporting threshold of $\pastro > 0.1$ for the subthreshold candidates in our full-bank search \cite{O2CatalogIAS}; it was listed as a subthreshold candidate in Ref.~\cite{NitzCatalog}. 
We determined the significance of GWC170620 in our targeted search by employing the data analysis techniques we have previously developed~\cite{pipelinepaper, ZackayPSDDriftPaper, templatebankpaper,ZackayFishing}. In addition, we develop an efficient ranking statistic that integrates the likelihood over the entire phase space of extrinsic parameters (including the Morse phase shift; presented in Appendix \ref{ap:ThirdImageStatistic}).

We determine that there is a probability $\approx 1.1 \%$ for the O2 BBH catalog to have a pair of events that are coincident at the level of GW170814 and GW170104. We compute this probability in a very conservative manner, by only relying on the coincidence between extrinsic (geometrical) parameters such as sky locations and Morse phases, and choosing to not interpret the matches between intrinsic parameters (this is akin to not using the factor of $\mathcal{B}^{\rm int}_{L/U}$ in Ref.~\cite{ConnorOGCLensingSearch}). 
Furthermore, we report a false alarm probability of ${}\approx 8 \times 10^{-3}$ that the targeted search uncovers a random noise event at the level of GWC170620.
However, as we will elaborate later, the measured relative Morse phases require a complex lens model with peculiar properties. This casts doubt on the lensing interpretation, but we do not have a better explanation for the intriguing parameter coincidences in the triplet.

We organize the rest of the paper as follows: in Section \ref{sec:methodology}, we explain the methodology of  determining the significance of the lensing association.
In Section \ref{sec:PE}, we perform joint parameter estimation of the three events under the lensing hypothesis. 
In Section \ref{sec:LensProperties}, we discuss the possible lens and source properties as inferred from the GW signals. 
We show that any viable lens must be a massive galaxy cluster, a possibility already considered for O1 and O2 BBHs in previous studies~\cite{2018MNRAS.475.3823S, 2019MNRAS.485.5180S}. 
We point out two potential candidate cluster lenses which can be investigated to look for a host galaxy. 
In Section \ref{sec:Discussion}, we give concluding remarks and discuss the prospects of finding the host galaxy. In Appendix \ref{ap:SignificancePair}, we explain in details how the significance of the candidate lensed events is quantified. In Appendix \ref{ap:LensingExpectations}
, we provide theoretical estimates on the occurrence probability of strong lensing events that resemble our candidate case. 

\begin{table*}
    \centering
    \caption{Summary of derived significance and inferred physical parameters. Time delays, Morse phase differences, and magnification ratios are quoted for the three events GW170104, GWC170620 and GW170814 ordered according to the event date.}
    \begin{tabular}{l|c|l}
        \tabline
        \tabline
        Item & Value & Reference\\
        \tabline
        Catalog FAP (GW170104, GW170814)& $1.1\times 10^{-2}$ & Section \ref{sec:methodology}\\
        Existence of GWC170620 (GPS time: 1181956460) & $1.3 \times 10^{-2}$ & Section \ref{sec:methodology}\\
        Time delays (relative to GW170104) & 0, 166.63 days, 222.01 days & Section \ref{sec:LensProperties}\\
        Morse phase differences (relative to GW170104) & 0, $\pi$, $\pi$ & Section \ref{sec:PE}\\
        Magnification ratios (relative to GW170814) & $0.401\pm 0.08$, $0.0719\pm 0.0024$, $1$ & Section \ref{sec:PE}\\
        Apparent luminosity distance of GW170814& $D_L^{\rm GW170814}/\sqrt{\mu_{\rm GW170814}} = 577^{+159}_{-216}\,{\rm Mpc}$ & Section \ref{sec:PE}\\
        Expected number of lensed events in O2 & $10^{-2}$--$10^{-3}$ & \cite{DaiGWLensingPopulation, NgPreciseLIGOLensingPredictions, Li:2018prc, OguriGWLensing, RobertsonLensingPredictions, ContigianiGWLensingEfficiency}\footnote{Note that cluster-scale lenses are neglected from the calculation in the references we list here, except for the simulation-based study of Ref.~\citep{RobertsonLensingPredictions}. We show in Appendix \ref{ap:LensingExpectations} that for low redshift sources dark matter halos significantly contribute to the lensing probability and alter the distribution of time delays.}\\
        \tabline
        
    \end{tabular}
    \caption*{
    As it can be seen, the system presents a challenge in any interpretation. On one hand, if it is not associated, explaining the fantastic association of GW170814, GW170104, along with the subthreshold trigger (GWC170620) requires a considerable amount of fine tuning.
    On the other hand, the expected lensing rate is small, as estimated from the redshift using standard assumptions in available literature.
    Moreover, the image types, order and magnification ratios require considerable fine tuning to produce, making the lensing interpretation less plausible.
    
    Whenever $\pm$ is used, one $\sigma$ error-bars are quoted. Whenever the $A^{+B}_{-C}$ notation is used, these are 95\% confidence intervals.}
    \label{tab:SystemIfact}
\end{table*}

\section{Determining the significance of the physical association}
\label{sec:methodology}

We begin by identifying pairs of events that are compatible with having either the same or related intrinsic and extrinsic parameters and weighting their significance.
This significance depends on many factors: (1) event SNRs; (2) time delays, strain amplitude ratios, and inferred orbital phase differences between events; (3) (detector-frame) chirp masses and spin parameters; (4) inferred source redshifts; (5) detectors that were operational at the times of the events. 

Each one of those may change the relative likelihood between the lensed ($\mathcal{H}_1$) and unlensed ($\mathcal{H}_0$) hypotheses by orders of magnitude.
Moreover, the inference also depends strongly on theoretical priors.
We aim to robustly quantify the significance of association in a way that is conservative in terms of prior choices.

In order to make inference independently of prior knowledge about intrinsic parameters, we first identify pairs of events that have consistent intrinsic parameters, and from there on we quantify the false alarm probability of the pair of events having coincident extrinsic parameters, for which prior distributions are known.

We quantify consistency in intrinsic parameters using the score 
\begin{align}
S_{\rm int} = \frac{\int\,\mathcal{L}_1(\theta)\,\mathcal{L}_2(\theta)\,\pi(\theta)\,\rmd\theta}{\left( \int\,\mathcal{L}_1(\theta_1)\,\pi(\theta_1)\,\rmd\theta_1 \right)\,\left( \int\,\mathcal{L}_2(\theta_2)\,\pi(\theta_2)\,\rmd\theta_2 \right)},
\end{align}
where $\theta$ stands for collectively the set of intrinsic parameters, and $\mathcal{L}$ is the likelihood, with a subscript denoting the event being used. The likelihood has been marginalized over all extrinsic parameters, which we collectively denote as $\psi$, with natural priors for angles, uniform priors for arrival times, and an Euclidean volumetric prior for the apparent source distance.
We then calculate the distribution of $S_{\rm int}$ given simulated pairs of events with the same parameters, from which we find a threshold score $S_{\rm min}$ such that $P(S_{\rm int}<S_{\rm min}|\mathcal{H}_1) < p_{\rm FN}$, where $p_{\rm FN}$ is the false negative probability of dismissing a genuine lensed image pair as unrelated.
Since the a priori probability for strong lensing is very small, and we seek only convincing cases, $p_{\rm FN} = 0.1$ is a reasonable choice.
Event pairs with $S_{\rm int} > S_{\rm min}$ are considered for further evaluation of extrinsic parameter coincidence.

Since the number of candidates passing the above selection is strongly mass dependent, we find it appropriate to differentiate BBH events by detector-frame chirp mass. For example, nearly all BBH event pairs with the chirp mass above $40M_\odot$ appear to have consistent intrinsic parameters, while for BBH events of lower chirp mass only two pairs out of a total of 15 in O2 pass the selection. Therefore, we estimate (admittedly subject to large uncertainty) the expected number of consistent pairs in O2 to be
\begin{equation}
    \langle N_{\rm consistent\;pairs}\rangle = 2
\end{equation}
As a sanity check, in Appendix \ref{ap:SignificancePair} Fig.~\ref{fig:Bint}, we also quantify the probability of finding a pair of events with intrinsic parameters appearing coincident at a level comparable to that of GW170104 and GW170814, under the assumption that the underlying BBH population having equal component masses and follow prior distributions of isotropic component spin vectors and flat chirp mass.
Under this assumption, the occurrence probability for a pair of events randomly having as good agreement in intrinsic parameters is $1.8\% \ll 2/15$.

Next we examine the more model-independent information from extrinsic parameters. To that end, we define a score:
\begin{widetext}
\begin{align}
S_{\rm ext} = \frac{\sum_{\phi_M}\,\int\,\rmd\psi\,\rmd(A_1/A_2)\,\mathcal{L}_1(\psi)\,\mathcal{L}_2(\psi,\phi_M,A_1/A_2)\,\pi(\phi_M, A_1/A_2, \Delta t)\,\pi(\psi)}{P(\Delta t | \mathcal{H}_0)\int\,\rmd\psi_1\,\mathcal{L}_1(\psi_1)\,\pi(\psi_1)\,\int\,\rmd\psi_2\,{\mathcal{L}_2(\psi_2)\,\pi(\psi_2)}}
\end{align}
\end{widetext}

Here $\psi_1$ and $\psi_2$ stand for the full lists of extrinsic parameters separately describing the two events, respectively, $\psi$ is the set of extrinsic parameters common to the two events under the hypothesis of multiple images, excluding the relative Morse phase $\phi_M$, the amplitude ratio $A_1/A_2$, and the arrival time difference $\Delta t$. Summation is performed over all possible values $\phi_M=0,\,\pi/2,\,\pi,\,3\pi/2$ allowed in geometrical lensing.

Since the value of $S_{\rm ext}$ is prior-dependent, for meaningful interpretation of it we compare the derived value to the $S_{\rm ext}$ distribution derived from simulated event pairs of chance coincidence with the same $S/N$ and detector sensitivity combination.
This renders its absolute normalization irrelevant.
Furthermore, we dramatically simplify the computation by noticing that the priors change more slowly than the likelihood does and hence priors flat in $\phi_M$, $\log(A_1/A_2)$ and $\Delta t$ can be used, as their actual values at their measured value could be taken into account later in the computation.
Details of computing $S_{\rm ext}$ are discussed in Appendix \ref{ap:SignificancePair}.

Having all the components in place, we quantify the probability for a catalog just like what we have to produce by chance a pair of events with consistent intrinsic parameters and equal or better agreement in extrinsic parameters:
\begin{equation}
\label{eq:FAP170104170814}
\begin{split}
&P({\rm better\:catalog}) \\
&\qquad\qquad= N_{\rm consistent\: pairs}\,P(S_{\rm ext}>S_{\rm ext}(\rm pair))\\
&\qquad\qquad= 1.1\times 10^{-2}.
\end{split}
\end{equation}
As can be seen, the low probability derived for the pair GW170814 and GW170104 warrants further consideration.
From just this pair of events, however, no additional information is accessible, while the theoretical lensing probability is very low (see Appendix \ref{ap:LensingExpectations}). Therefore, information gathered thus far does not yet constitute sufficient evidence to claim a candidate.

In the search effort of Ref.~\cite{HannukselaGWLensing} to find lensed multiple images in the catalog of LVC events, the pair GW170104 and GW170814 were brought up as one of the most interesting pairs, but were not assigned a high overall significance due to theoretical prior against long time delays. When neglecting the suppression of the overall significance from the theoretical prior on the time delay, however, an intriguingly high Bayes factor $\sim 250$ was reported favoring common source parameters (except for the orbital phase, distance and times) over unrelated ones. Our overall significance of parameter coincidence Eq.~(\ref{eq:FAP170104170814}) is lower than that number because we have chosen to exclude the inference from consistent intrinsic parameters. We have made this choice because we prefer to report a conservative significance insensitive to theoretical priors on BBH mass and spin distributions, and because we worry about potential numerical artifacts that could bias the results when carrying out Bayes integration over the entire high-dimensional parameter space. Moreover, we note that the search for additional lensed images performed by Ref.~\cite{ConnorOGCLensingSearch} also recovered the pair GW170814 and GW170104 as having the best waveform match, although the authors did not further consider them for the lensing hypothesis based on the conclusion of Ref.~\cite{HannukselaGWLensing}.

If the association is genuine, as we will discuss later in Section \ref{sec:LensProperties}, the long time delays between images point to moderate magnification factors, a low source redshift, and a high internal velocity dispersion for the lens. This hint of a massive, complex lens motivates us to search for additional lensed images in the data.

The association of GW170814 and GW170104 informs us of a well determined source sky location, a single waveform template and partial information about the inclination and polarization angle. This should dramatically reduce the search space and improve the search sensitivity for uncovering additional images with lower SNR.
To exploit that, we implement a coherent score as a ranking statistics that takes into account all possible information we have collected from the pair. Details are presented in Appendix \ref{ap:ThirdImageStatistic}.
This targeted search achieves a factor of $\sim 10^4$ reduction in the look-elsewhere effect compared to our standard search, making it $\sim 20\%$ more sensitive in strain amplitude.

This targeted search for sub-threshold signals uncovers GWC170620, a candidate for an additional lensed image associated with GW170814 and GW170104. 
The statistical significance of this third candidate image is determined using time slides. We derive a false alarm probability $1.3 \times 10^{-2}$ for a random background signal to have an equal or higher coherent score than what is found for GWC170620 (see Appendix \ref{ap:ThirdImageStatistic}). The interpretation that GWC170620 is a genuine but unrelated BBH event is substantially less likely.

The discovery of GWC170620 strongly augments the significance of the coincidence found between GW170814 and GW170104. The probability of all this taking place purely by chance is as small as $\sim 10^{-4}$.

We note in passing that GW170202, an event uncovered in our full-bank search~\cite{O2CatalogIAS}, also has consistent sky location and a Morse phase with the candidate triplet of GW170104, GWC170620 and GW170814.
Its intrinsic parameters are very close to the ones inferred for the triplet, but enforcing them to be exactly the same as the ones of the triplet causes significant reduction in the likelihood.
We therefore deem its association with the triplet as a fourth candidate image a weak case that certainly requires additional information.

We further mention that there is another set of signals consistent with being a lensed triplet, GW170425, GW170727, and GWC170321 (GPS time 1174138338.385). We measure the false alarm probability for a catalog to contain these signals by chance to be substantially higher, about 1\%. This is mainly due to larger measurement uncertainties in all source parameters, and the lack of a member signal with three-detector localization which would otherwise enable a tight measurement of the Morse phase differences. Note though that this system has a higher prior probability of being lensed as its inferred source redshift is substantially higher, and hence a dramatically larger lensing probability a priori (see Appendix \ref{ap:LensingExpectations}). 
While this set of three signals are also consistent with being lensed, without additional information we do not think this is an equally strong case as the set of GW170104, GWC170620 and GW170814.

\section{Joint parameter estimation}
\label{sec:PE}

Under the lensing hypothesis, we perform parameter estimation using all three BBH signals. We fit the data to the \texttt{IMRPhenomD} waveform model~\cite{Khan:2015jqa} under the assumption that both component spin vectors are (anti-)aligned with the orbital angular momentum vector. There are four source intrinsic parameters common to all three events, namely the two component masses $m_1$ and $m_2$ (in the detector frame) and the dimensionless spin magnitudes $\chi_{1z}$ and $\chi_{2z}$ perpendicular to the orbital plane. We adopt the same mass prior as in Ref.~\cite{O2CatalogIAS} and a prior flat in the effective spin parameter $\chi_{\rm eff}=(m_1\,\chi_{1z} + m_2\,\chi_{2z})/(m_1+m_2)$ as defined in Ref.~\cite{GW151216}.

Lensed image candidates share the same right ascension and declination $(\alpha,\,\delta)$ for source sky location, orbital inclination $\iota$, and polarization angle $\psi$, all under natural priors. Since lensing time delays and magnification ratios are unknown, for each event we independently fit for an arrival time and a strain amplitude normalization. The amplitude normalizations are parameterized in terms of the apparent luminosity distance $D_L$ of GW170814 with a prior uniform in the Euclidean volume $P(D_L) \propto D^2_L$ in the range $0 < D_L < 10\,$Gpc, and flux magnification ratios of GW170104 and GWC170620 relative to GW170814 with log-flat priors.

As explained before, the inferred orbital phases $\varphi$ from lensed images should have values that differ by integer multiples of $\pi/4$. Aiming to test the data against the lensing hypothesis, we intentionally fit each event to an independent orbital phase with a flat prior within the range $[0,\,\pi)$, as appropriate for the dominant $(\ell,\,m)=(2,\,2)$ radiation harmonic. We checked that the measured orbital phase differences are uncorrelated with other parameters. Hence, inference results for other parameters are approximately unaffected by our prior choice for the orbital phases. 

\begin{figure}
     \centering
     \includegraphics[width=\linewidth]{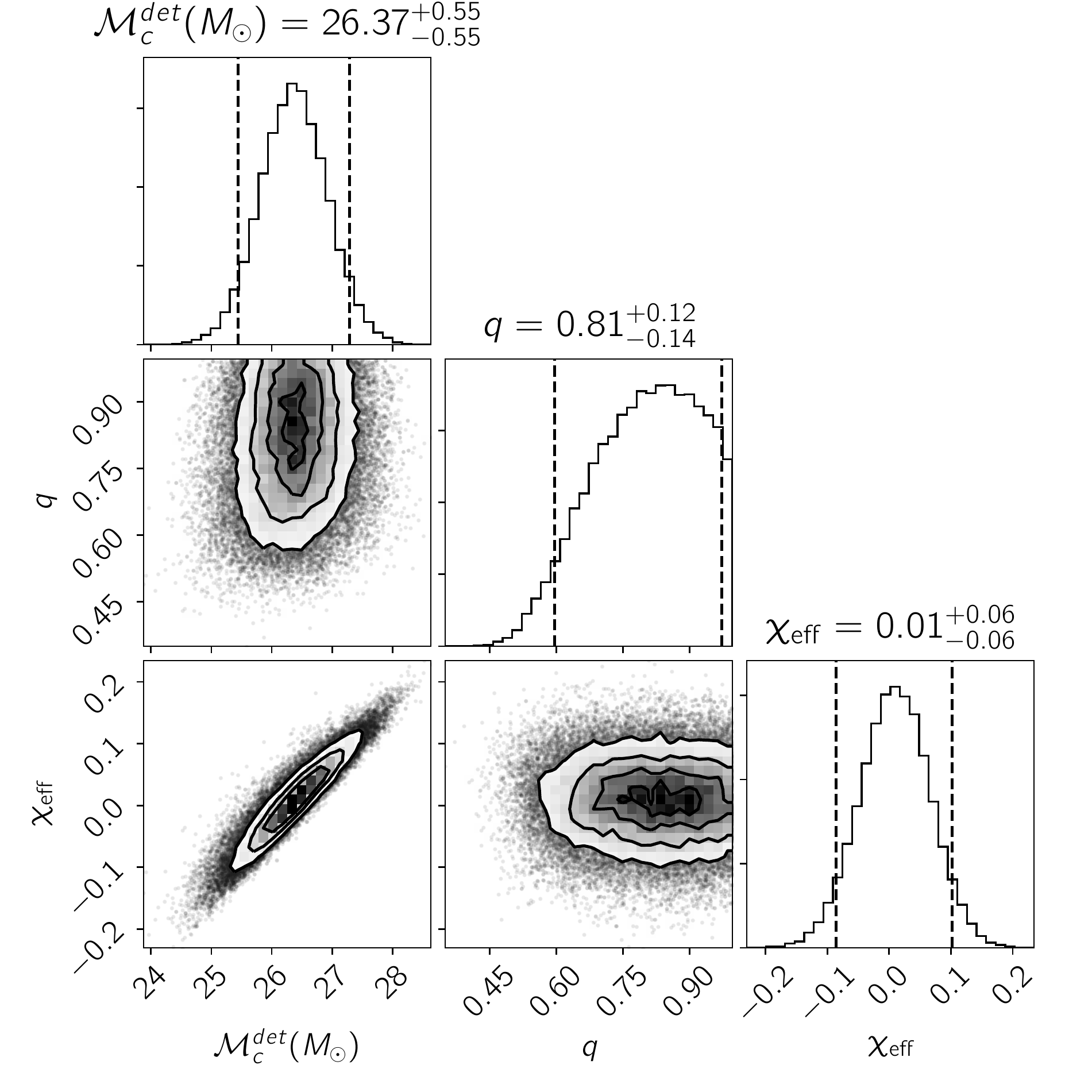}
     \caption{Posterior distributions for the detector-frame chirp mass $\mathcal{M}_c^{\rm det}$, mass ratio $q=m_2/m_1$, and the effective spin parameter $\chi_{\rm eff}$ as inferred from a joint fit of GW170104, GW170814 and GWC170620. Contours enclose area containing 39\% and 86\% of the probability. Error bars indicate the 95\% and 5\% percentiles.}
     \label{fig:int_parameters}
 \end{figure}

We use \texttt{PyMultiNest}~\cite{PyMultiNest} for sampling the posterior, and compute the likelihood using the technique of relative binning~\cite{Zackay2018}.
The posterior for the intrinsic parameters is shown in Fig.~\ref{fig:int_parameters}. The inferred parameters remain consistent with the ones reported in \cite{GWTC-1}, but with smaller uncertainties.

\begin{figure}
     \centering
     \includegraphics[width=\linewidth]{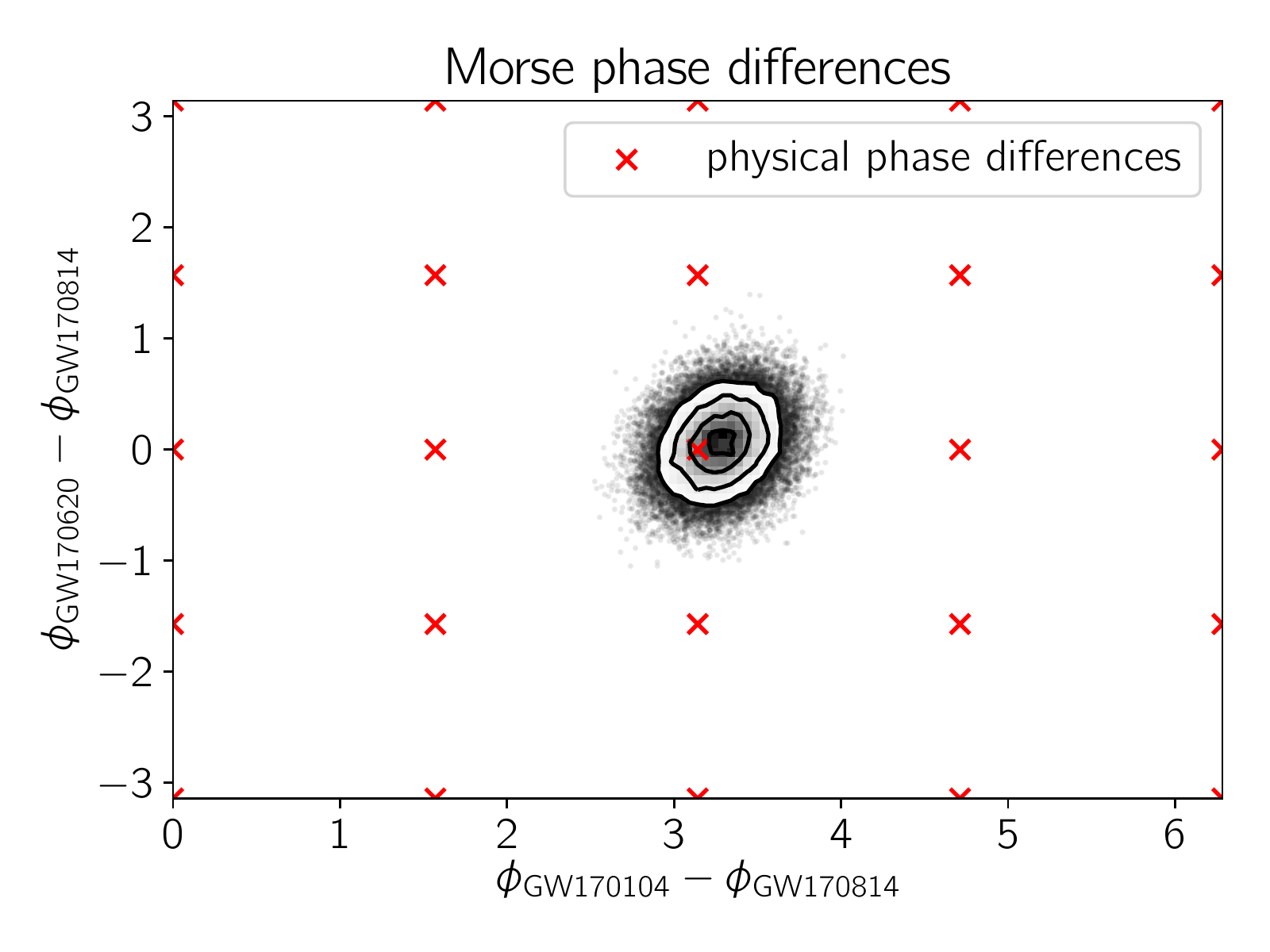}
     \caption{Posterior distribution for the Morse phase differences as inferred from a joint fit of GW170104, GW170814 and GWC170620. The phase space has period $2\pi$ along both dimensions. Marked with red crosses are all combinations of values physically possible with geometric lensing. Contours indicate 39\%, 86\% and 98\% of the probability.}
     \label{fig:morse_phase}
 \end{figure}

Figure~\ref{fig:morse_phase} shows measured differences in the Morse phases $\phi$ as converted from differences in the orbital phase via $\Delta\phi = 2\,\Delta\varphi$ (with 1$\sigma$ uncertainties):
\begin{align}
 & \phi_{\rm GW170104} - \phi_{\rm GW170814} = 3.28^{+0.18}_{-0.18}\,\,{\rm rad}, \\
 & \phi_{\rm GWC170620} - \phi_{\rm GW170814} = 0.06^{+0.27}_{-0.27}\,\,{\rm rad}.
\end{align}
The Morse phase differences are strikingly consistent with the choice
$\phi_{\rm GW170104} - \phi_{\rm GW170814} = \pi$ and $\phi_{\rm GWC170620} - \phi_{\rm GW170814} = 0$. Uncertainties on these measurements are small enough to rule out any other choices allowed by geometric lensing. As we would like to emphasize, mutual orbital phase differences between a set of lensed image candidates are well measurable despite large uncertainty in the orbital phase of any individual event due to parameter degeneracy. For the three events under our scrutiny, this is curious evidence supporting lensing association. If these events are unrelated, measurements most likely would have yielded random values that are inconsistent with any of the allowed values as indicated by red crosses in Fig.~\ref{fig:morse_phase}. The lensing explanation, assuming all values indicated by the red crosses are equally probable, is $\sim 6$ times more likely than a random outcome agreeing with one of the red crosses by chance. 
 
\begin{figure}
     \centering
     \includegraphics[width=\linewidth]{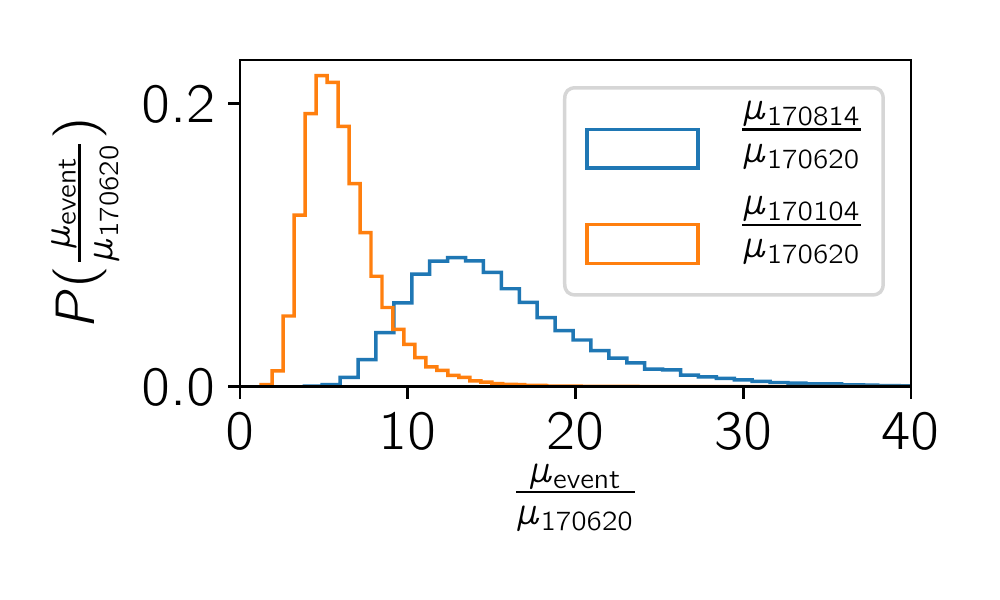}
     \caption{Inferred magnification ratios between lensed image candidates GW170814, GW170104 and GWC170620. The magnification ratios are determined solely from the GW data without imposing time-delay or image-topology priors.}
     \label{fig:mag_ratios}
 \end{figure}

Figure~\ref{fig:mag_ratios} shows the distribution of inferred magnification ratios between the three events (median and $68\%$ confidence interval):
\begin{align}
 & \mu_{\rm GW170814} / \mu_{\rm GWC170620} = 14.5^{+5.9}_{-3.9}, \\
 & \mu_{\rm GW170104} / \mu_{\rm GWC170620} = 5.7^{+2.4}_{-1.6}.
\end{align}
The measured large magnification ratios are impressive, owing to the facts that the SNR of GW detection is proportional to wave amplitude and not flux, and that detector response to a fixed source sky location varies significantly with the sidereal hour which enables fortuitous detection of faint lensed images.

\begin{figure}
     \centering
     \includegraphics[width=\linewidth]{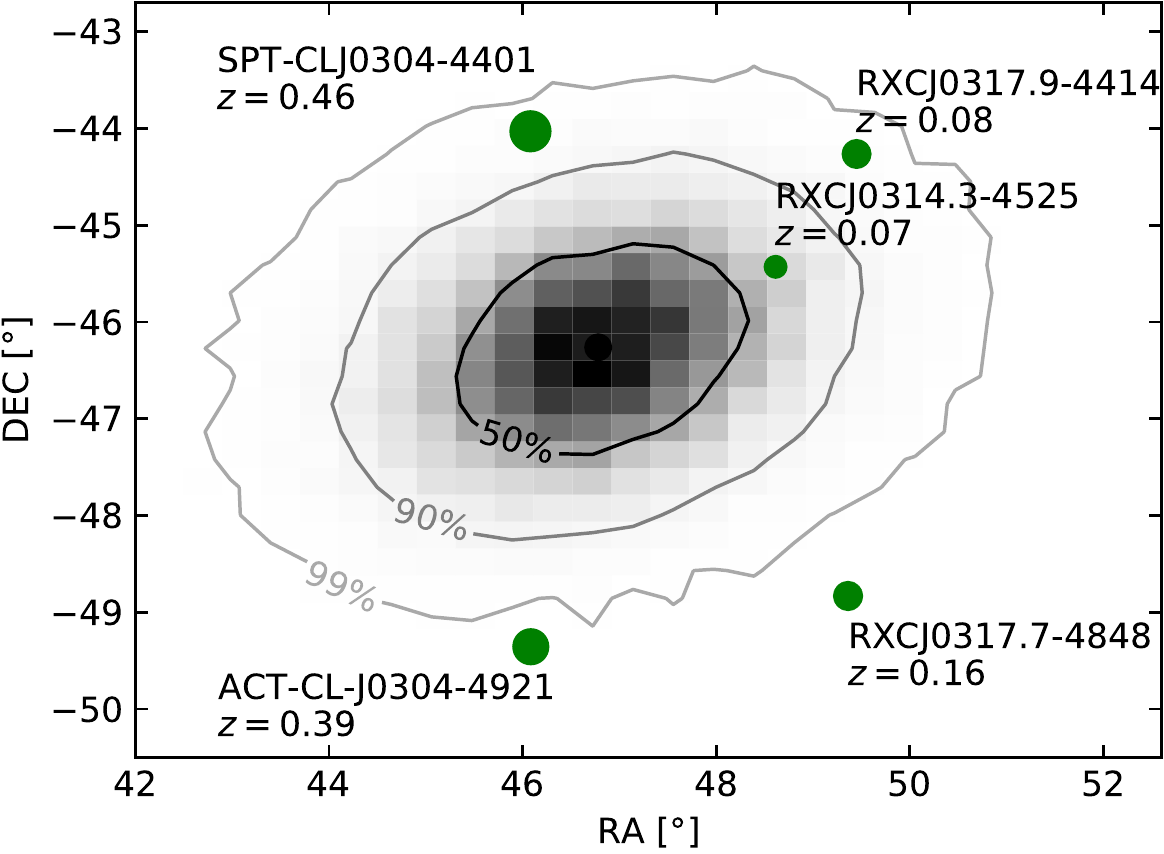}
     \caption{Source sky location inferred under the lensing hypothesis for the image candidates GW170814, GW170104 and GWC170620. Clusters found in the Planck catalog \cite{2016A&A...594A..27P} are shown as green disks, whose relative sizes indicate cluster mass.}
     \label{fig:sky_localization}
 \end{figure}

As shown in Fig.~\ref{fig:sky_localization}, our joint parameter inference localizes the source to within an error ellipse of only $16\,{\rm deg}^2$ ($90\%$ confidence interval) on the sky. 

\section{Viable Lens Properties}
\label{sec:LensProperties}

Through GW parameter inference, we have measured the Morse phase differences, magnification ratios, and arrival time differences between the lensed image candidates.
The observed strain amplitudes constrain the true source luminosity distance divided by the square root of the lensing magnification, up to degeneracy with the orbital inclination.
These pieces of information strongly constrain the viable lens.

Under the approximation of a single lens plane, the measured Morse phase differences imply that GW170104, GWC170620 and GW170814, ordered by their arrival times, have image types $\rm L,\,H,\,H$ or $\rm H,\,L,\,L$, where L/S/H stand for minimum/saddle/maximum points of the Fermat potential, respectively, following the notation used by Ref.~\cite{1986ApJ...310..568B}.

Provided that there are no discontinuities in the Fermat potential, the number of images of various types obey the following equation~\cite{1986ApJ...310..568B},
\begin{align}
    n(L) + n(H) - n(S) = 1\,.
\end{align}
Therefore, at least two S images must have been missed.
In the ``HLL'' scenario, there must have been at least another L image, i.e. the global minimum of the Fermat potential, that arrived earlier than GW170104. In this case, there must have been at least 4 missed images out of a total of 7 images. The fraction of time during O2 when two or more detectors were simultaneously observing was less than $50\%$. Moreover, the detector antenna patterns also present order unity variation as the Earth rotates relative to the sidereal sky. Hence it may not be surprising if about half of the images had evaded detection.

Since L images always have magnification factors greater than unity~\cite{1984A&A...140..119S, 1986ApJ...310..568B}~\footnote{There can be possible violations of this statement in an expanding universe because the lensing convergence may be negative if part of the lens has a mass density lower than the cosmic mean density. In practical situations of galaxy or cluster lensing, this violation is unlikely.}, a lower limit for the source redshift $z_S$ can be derived from the faintest L image (using the 5\% percentile of the measured GW amplitude):
\begin{align}
    z_S^{\rm HLL} > 0.26\,, \\ 
    z_S^{\rm LHH} > 0.13\,.
\end{align}
However, both scenarios require a magnified H image. As far as we are aware of, this is uncommon in galaxy or cluster lensing of quasars and supernovae. In the ``HLL'' scenario, the last image GW170814 is the brightest of all detected images. In the ``HLL'' scenario, the intermediate {\rm L} image, which is the faintest detected image candidate, must have a magnification larger than unity and thus the first {\rm H} image has a magnification greater than about 6. Although in principle the observed Morse phase differences can be accommodated by a physical lens, it is not one typical of what have been observed for lensed quasars and galaxies.

Absolute magnification factors are not measurable from GW data alone. Large magnifications typically arise when the source is close to a lensing caustic. Had the magnification of GW170814 been very high, there should have been another image that is almost equally magnified but with opposite parity, i.e. an S image. For a crude estimate, the time delay between a close pair of highly magnified images should be of the order
\begin{equation}
\begin{split}
        \Delta t
        & \sim (1+z_L)\,\frac{D_L\,D_S}{c\,D_{LS}}\,
            \frac{\Delta\theta^3}{\theta_L} \\
        &\sim (1+z_L)\,\frac{D_L\,D_S}{c\,D_{LS}}\,\frac{\theta^2_L}{\mu^3} \\
        & \sim \SI{0.02}{d}\, (1+z_L)
            \left( \frac{D_L\,D_S / D_{LS}}{\SI{400}{Mpc}} \right)
            \left(\frac{\theta_L}{10''}\right)^2 \left(\frac{40}{\mu}\right)^3.
    \end{split}
\end{equation}
Here $\Delta\theta$ is the image separation, $\theta_L$ is some characteristic angular scale of the lens, and $\mu$ is the magnification factor of the image pair. We introduce the lens redshift $z_L$, and the angular diameter distances to the lens $D_L$, to the source $D_S$, and from the lens to the source $D_{LS}$, whose fiducial values are taken to be $z_L=0.08$ and $z_S=0.5$. 

With an intraday time delay, how could a bright counter image of GW170814 have evaded detection? It is worthy to note that even if detectors remain stably operational over many hours around the times of close image multiplets, a bright counter image might still be undetectable to the LIGO detectors due to the twice-daily variation of the detector antenna patterns on the sidereal sky~\cite{BroadhurstSmootGWLensingCandidate}.

\begin{figure}
     \centering
     \includegraphics[width=\linewidth]{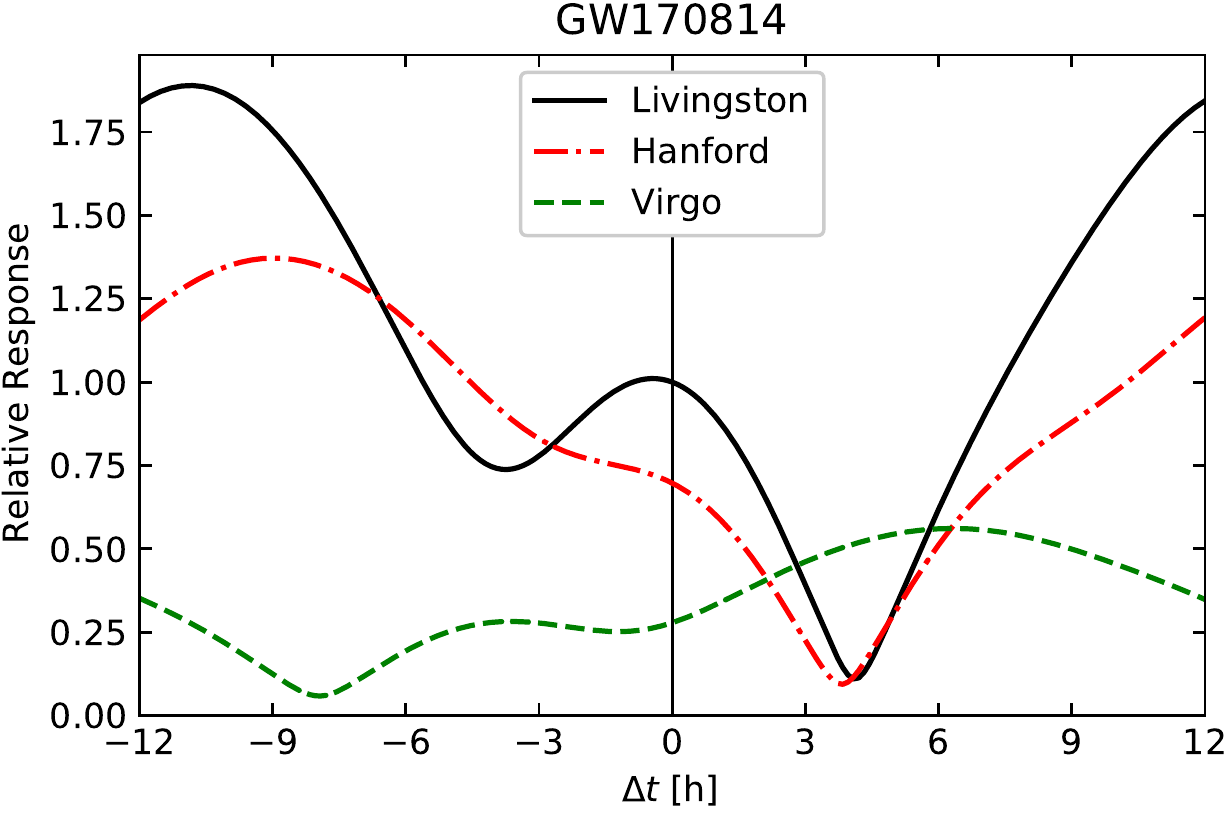}
     \caption{Strain amplitude responses for the Livingston, Hanford and Virgo detectors around the time of GW170814, computed for a face-on source at the sky location of GW170814 and normalized to the Livingston response at the time of the event.}
     \label{fig:rel_resp}
 \end{figure}

We find out that, if GW170814 has a close counter image of equal magnification, that image could have evaded detection if it arrived $\sim 4$--$5$ hours after GW170814 (viable for the ``HLL'' scenario) when both LIGO detectors became almost blind toward the inferred sky location. During that period of time, the Virgo detector happened to have a favorable antenna response and may have recorded a significant signal. This can be seen in \reffig{rel_resp} in terms of the detector strain responses to a face-on binary source. We therefore carry out a single-template, directional search for coherent signals in the LIGO--Virgo network targeting from 2 hours before to 8 hours after the time of GW170814. An intriguing signal consistent with the Morse phase of an S image and equal magnification ratio with GW170814 is recovered $3.8$ hours after GW170814 at a network SNR of 5.5. Unfortunately, this signal is indistinguishable from Gaussian random noise according to our false alarm rate calculation. Our analysis nevertheless suggests that additional equally bright images might have been hidden in noise during normal LIGO/Virgo operation times due to antenna pattern variations. In the regime of high magnification, exactly equal image pair might also be idealistic as the flux becomes prone to perturbation from small-scale substructure lenses~\citep{Diego:2019lcd, Dai:2018mxx, Dai:2020rio}, which also makes room for the possibility of missed images with short time delays.

As we will justify shortly, the probable Einstein scale would be that of a galaxy cluster lens $\theta_L \sim 10''$. For sufficiently high magnification $\mu \gtrsim 40$, the expected time delay between the highly magnified image pair becomes short enough $< \SI{1}{h}$ that detector response variation becomes irrelevant. Absence of such a counter image within $\SI{1}{h}$ following GW170814, according to our targeted search, implies that:
\begin{equation}
    z_S \lesssim 0.7\,.
\end{equation}
This suggests that the source and hence the lens should have relatively low redshifts, contrary to the typically observed situations of strongly lensed background galaxies or quasars.

For a singular isothermal lens, the characteristic internal velocity dispersion can be estimated as
\begin{align}
    \sigma & \sim \frac{c}{\sqrt{4\pi}}\,\left(1+z_L\right)^{-1/4}\,\left(\frac{t_d\,c\,D_S}{D_L\,D_{LS}}\right)^{1/4} \nonumber\\
    & \sim \frac{400\,{\rm km/s}}{(1+z_L)^{1/4}}\,\left(\frac{t_d}{200\,{\rm d}}\right)^{1/4}\,\left( \frac{D_L\,D_{LS}/D_S}{260\,{\rm Mpc}} \right)^{-1/4}.
\end{align}
The long time delays between detected candidate images imply that the lens should be a massive cluster of galaxies. It is interesting to note that if the source redshift is relatively low $z_S \lesssim 0.7$, important contribution to the lensing cross section comes from intervening galaxy clusters with velocity dispersion $\SI{500}{\kilo\meter/\second}< \sigma < \SI{800}{\kilo\meter/\second}$ in addition to individual galaxy lenses.
See Appendix \ref{ap:LensingExpectations} for more details.

Since massive clusters are rare, we expect very few candidate lenses within the localization error ellipse shown in Fig.~\ref{fig:sky_localization}. Within that region, we have found two clusters, A3104 (RXCJ0314.3-4525) and A3112 (RXCJ0317.9-4414), with roughly the correct velocity dispersion, and low lens redshifts which are more compatible with the inferred low source redshift.

Under the assumption of a single lens plane, it seems inevitable in either the ``LHH'' or the ``HLL'' scenario that a magnified H image forms due to a remarkably and perhaps unrealistically shallow central profile in a component lens. In principle, this can be avoided if there are two lens planes. However, even in this case both lens planes need to be super-critical, a situation that seems extremely improbable to arise. 
 
Therefore, the lensing interpretation of the triplet GW170814, GW170104 and GWC170620 appears to require a complex lens. This was unexpected to be the first lens found in GW observation according to many theoretical calculations~\citep{Li:2018prc, OguriGWLensing}, although recent studies based on numerically simulated lens populations have highlighted the contribution of cluster-scale lenses for high magnification events~\citep{RobertsonLensingPredictions}. Follow-up electromagnetic study of these clusters and other candidate lenses would be of great value to confirm or falsify this interpretation.

\section{Conclusion}
\label{sec:Discussion}

We have performed a search for multiple lensed images of the same GW sources in LIGO O2, incorporating information from the Morse phases for the first time. We have identified a set of three candidate lensed images of GWs with long time delays that may have originated from the same BBH merger.
The association rests on agreement in intrinsic and extrinsic parameters of two catalogued events GW170814 and GW170104, and a subthreshold candidate GWC170620, with a chance probability ${}\approx 10^{-4}$.
While this chance probability is low, the a priori probability of detecting multiple lensed images of the observed type during O2 is even lower. We therefore judge that these candidates do not yet make a confident case of strong lensing.

If these are genuine lensed events, the long time delays imply that the lens is likely to be a cluster or a group of galaxies. However, the image types as inferred from the Morse phases, together with magnification ratios and time delays, point to a peculiar image configuration, casting doubt on the lensing interpretation.

We narrow down the sky location of the viable source to within $\sim 16\, {\rm deg}^2$, presenting the astronomy community with a rare opportunity to identify the host galaxy of a BBH merger. Any lens-host candidate system would be put under stringent test based on the observed time delays and magnification ratios. 

Host galaxies of long gamma ray bursts are known to be often intrinsically faint and small star-forming galaxies~\citep{savaglio2009galaxy}. If the typical host galaxies of BBH mergers share similar properties, especially for heavy BBHs~\citep{elbert2018counting}, it may be challenging to detect them from cosmological distances even with some amount of magnification. The intrinsic optical luminosity of the host galaxies sensitively depend on the BBH merger delay timescale, because a long delay timescale would mean shutdown of star formation at the time of observation and hence a fainter galaxy. In the case examined here, however, the expected low source redshift suggests that the prospect of finding the host galaxy in deep surveys of lensed galaxies might be more optimistic than on average.

BBH coalescence may take place far away from a host galaxy if BBHs are subject to strong natal kicks and if the delay timescale to merger is long after binary formation. In this case, GWs from the BBH merger and the host galaxy may be magnified by very different amounts. Currently, this scenario of large source-host separation lacks empirical evidence~\citep{Mandel:2015eta}.

As inferred from the image types, several lensed images have been missed (at least 2 for ``LHH'' and 4 for ``HLL''). We therefore encourage the examination of any additional usable data that the LIGO--Virgo collaboration might have during engineering runs when the interferometers were locked. Any additional associated event will further constrain the lens and make the lensing case definitive.

If a host galaxy is found, one will be able to derive the magnifications, image types, and approximate arrival dates for the other images. Moreover, accurate knowledge of the time delays will enable to constrain source location within the host, providing valuable insight into the formation mechanism of BBHs.

\section*{Acknowledgments}

The authors thank Parameswaran Ajith, Keith Bechtol, Jo Dunkley, Alexander Kaurov, Sigurd Naess, Masamune Oguri, Eli Waxman and Adi Zitrin for helpful discussions. We greatly thank the LIGO Collaboration and the Virgo Collaboration for making the O1 and O2 data publicly accessible and easily usable.

This research has made use of data, software and/or web tools obtained from the Gravitational Wave Open Science Center (\url{https://www.gw-openscience.org}), a service of LIGO Laboratory, the LIGO Scientific Collaboration and the Virgo Collaboration. LIGO is funded by the U.S. National Science Foundation. Virgo is funded by the French Centre National de Recherche Scientifique (CNRS), the Italian Istituto Nazionale della Fisica Nucleare (INFN) and the Dutch Nikhef, with contributions by Polish and Hungarian institutes.

LD and TV acknowledge support from the John Bahcall fellowship. LD acknowledges support from the Raymond and Beverly Sackler Foundation Fund. BZ acknowledges the support of the Peter Svennilson Membership Fund and the Frank and Peggy Taplin membership fund. TV acknowledges support by the Friends of the Institute for Advanced Study. TV also thanks the International Centre for Theoretical Sciences (ICTS) for hosting him during the period in which this manuscript was written. MZ is supported by NSF grants PHY-1820775 the Canadian Institute for Advanced Research (CIFAR) Program on Gravity and the Extreme Universe and the Simons Foundation Modern Inflationary Cosmology initiative.

\appendix

\section{Significance of the GW pair}
\label{ap:SignificancePair}

\subsection{Intrinsic parameter agreement}

Assessing the significance of parameter coincidence between candidate lensed events is inherently uncertain to some degree. This is mainly because it depends on our theoretical prior on BBH mass and spin distributions. The agnostic space to compare the agreement between intrinsic parameters of two events is arguably the space of coefficients $c_\alpha$'s of orthonormal base phase functions using the formalism of geometric template placement we lay out in Ref.~\cite{templatebankpaper}. In this space, the likelihood function is well approximated by a multi-variate Gaussian, with a covariance matrix that is proportional to the identity matrix and scales with the inverse of the squared SNR. This is exactly true if the PSD used to construct the template bank coincides with the one at the event time. In practice, deviation from that causes negligible corrections. 

\begin{figure*}
     \centering
     \includegraphics[width=\linewidth]{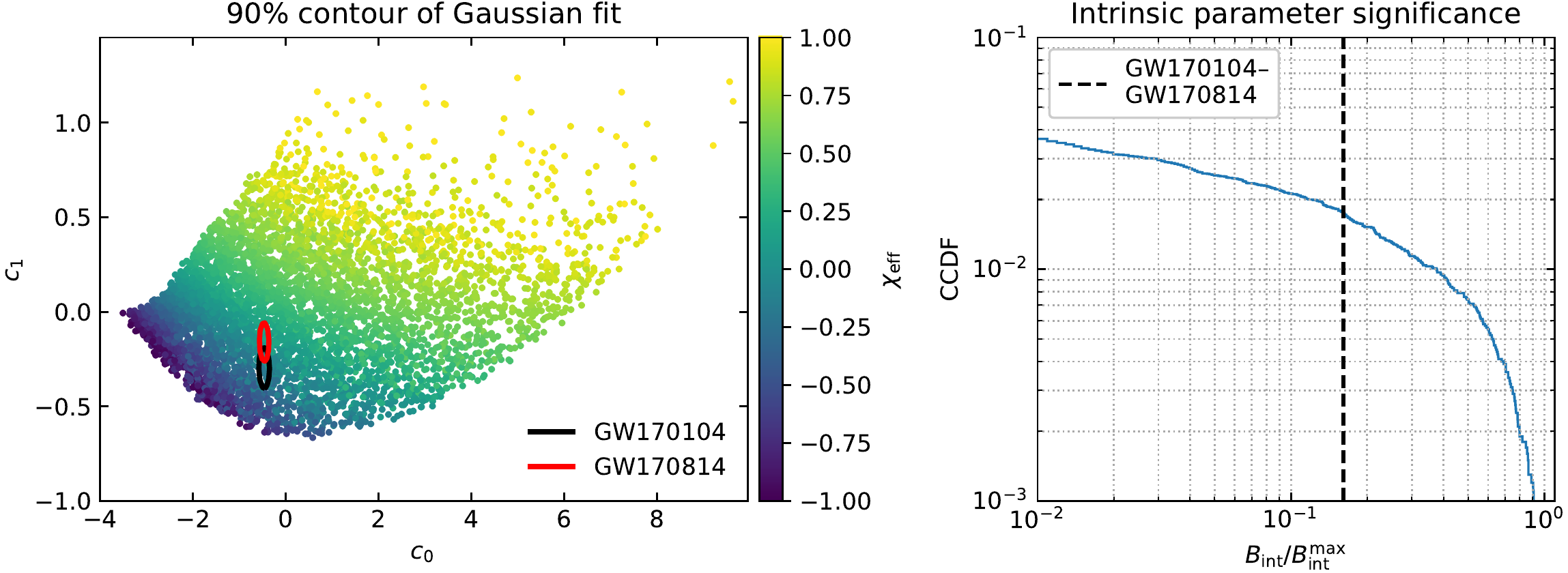}
     \caption{Significance of coincidence in intrinsic parameters for GW170814 and GW170104. {\it Left:} We assess coincidence in intrinsic parameters in the $c_\alpha$ space. We project randomly generated physical waveforms into this space and color-code them according to the value of $\chi_{\rm eff}$.
     In this space, the posteriors of GW170814 and GW170104 have small widths compared to the extent of the physical region and overlap with each other. If restricted to values close to $\chi_{\rm eff}=0$, however, the chance probability of coincidence becomes much higher, which would lead to very different inference on the significance of $B_{\rm int}$.
     {\it Right}: Complementary cumulative distribution of $B_{\rm int}$ (Eq.~(\ref{eq:Bint})) under the null hypothesis with the value of $B_{\rm int}$ observed for GW170814 and GW170104 indicated by the vertical dashed line. The distribution is derived from waveform injections assuming a uniform chirp mass prior between $20\ M_\odot$ and $40\ M_\odot$, a uniform mass ratio prior between 0.7 and 1, and a spin prior for which the dimensionless spin vectors for binary components independently have magnitudes uniformly distributed between 0 and 1 and have random orientations. We note that the isotropic spin prior adopted here favors low $\chi_{\rm eff}$ values.
     }
     \label{fig:Bint}
 \end{figure*}

Given an event, we compute the incoherent squared SNR and fit the likelihood to a Gaussian in the $c_\alpha$ space. GW170814 and GW170104 belong to $\texttt{BBH (3,0)}$ bank which has only two large $c_\alpha$ dimensions. 
In the left plot of \reffig{Bint}, we show the contours of the fitted Gaussian for GW170814 and GW170104. The compactness of the contours can be compared to the overall extent of the region with physical templates, for which we show a set of randomly generated physical templates used to construct the template bank. The gradient in the value of $\chi_{\rm eff}$ is evident. The dimension perpendicular to that gradient approximately follows the change in the chirp mass. Clearly, the false alarm probability of having a pair of events with coincident waveforms as GW170814 and GW170104 is sensitive to the actual $\chi_{\rm eff}$ distribution of BBH sources.

To quantify the coincidence in intrinsic parameters, we inject waveforms according to a prior that is uniform in chirp mass between $20\ M_\odot$ and $40\ M_\odot$. To be conservative, we restrict the mass ratio to be uniformly distributed only between 0.7 and 1. We consider a spin prior that favors low $\chieff$ values by independently drawing random spin vectors for binary components, whose magnitudes are uniformly distributed between 0 and 1 and have random orientations. 

Due to noise, the maximum likelihood $c_\alpha$'s recovered from an injected signal are distributed around the injected values with a variance that is inversely proportional to the squared SNR. We verify and calibrate the relation between injected and recovered $c_\alpha$ values using a relatively small number of injections, and then take advantage of this relation to bypass having to inject a computationally prohibitive number of signals. 

Fixing the peak location of the Gaussian likelihood for the louder event GW170814, we draw random peak locations for the analog of GW170104 following the aforementioned trick. We then quantify the overlap between the two Gaussians by computing the following Bayesian evidence ratio:
\begin{equation}
\label{eq:Bint}
    B_{\rm int} = \frac{A_{\rm eff} \int \rmd^2 c\, G_1(c)\,G_2(c)}
                       {\left(\int \rmd^2 c\, G_1(c) \right)\, \left( \int \rmd^2 c\, G_2(c) \right)},
\end{equation}
where $G_i(c)$'s are Gaussian approximation of the likelihood functions for the pair of events, and we parameterize the prior probability for the peak location in $c_{\alpha}$-space as $1/A_{\rm eff}$ in the vicinity of the parameters of GW170814. There is a maximum value for $B_{\rm int}$, which we define as $B_{\rm int}^{\rm max}$, when both $G_i$'s coincide in the peak location. The right plot of Fig.~\ref{fig:Bint} shows that only 1.8\% of the injections have a higher score than what is found for GW170814 and GW170104 under our choice of prior.

\subsection{Extrinsic parameters}

We now assess the coincidence in extrinsic parameters between a pair of candidate lensed images. To stay computationally feasible, we will fix a set of intrinsic parameters that well fit both events. For the set of candidates under consideration here, GW170814 is recovered with the highest SNR and is well localized on the sky, so we will phrase the question as how consistent the extrinsic parameters of GW170104 are with those of GW170814 with intrinsic parameters fixed to be the best-fit of GW170814. In particular, we will use the maximum likelihood waveform of GW170814 in the $c_{\alpha}$-space. 

Among the full set of extrinsic parameters, the geocentric arrival times and inferred source distances of GW170814 and GW170104 provide little information because time delay and magnification ratio between lensed images are a priori unknown. Therefore, we consider if candidate lensed images have compatible sky location $({\rm RA}, {\rm DEC})$, source inclination $\iota$, polarization angle $\psi$, and orbital phase $\varphi$, with the Morse phase effect taken into account. 

We start with posterior samples obtained from full parameter estimation of GW170814. We disregard the intrinsic parameters and fix the waveform to have the maximum likelihood intrinsic parameters. One subtlety arises because the precise definition of the orbital phase $\varphi$ depends on the intrinsic parameters and the geocentric arrival time. Thus we recompute $\varphi$ for each set of other extrinsic parameters at fixed best-fit intrinsic parameters. 

For each sample set of parameters, we can compute the expected (complex-valued) overlap in the $k$th detector, $\bar Z_k$, between the underlying strain signal and the normalized template. For the dominant (2, 2) harmonic, it takes the following form 
\begin{equation}
    \bar Z_k = R_k\,f^{1/2}_k\,x_0\,e^{i\phi_0}
        \equiv T_k\,x_0\,e^{i\phi_0}
        \equiv T_k\,Y
\end{equation}
where $R_k$ is the complex detector response 
\begin{equation}
    R_k = F_{+,k}\,\frac{1+\mu^2}{2} - i\,F_{\times,k}\,\mu,
\end{equation}
where $\mu = \cos\iota$. The detector response coefficients $F_+$ and $F_\times$ are functions of RA, DEC and $\psi$~\cite{Sathyaprakash2009}. Different noise PSDs in different detectors are accounted by $f^{1/2}_k$. We have introduced the amplitude and phase constants of the signal, $x_0$ and $\phi_0$, respectively. In geometric lensing, once the phase of GW170814 is fixed, for a counterpart image $\phi_0$ can take only four possible values. On the other hand, the amplitude $x_0$ is unconstrained because magnification ratio is not known.

Consider the logarithm of the likelihood
\begin{equation}
  \ln L = -\frac 12\,\sum_k \left| Z_k - \bar Z_k \right|^2,
\end{equation}
where $Z_k$ is the complex-valued overlap between the strain data and the normalized template in detector $k$. Furthermore, we implicitly exploit the inferred sky location for GW170814 to predict the relative times to evaluate $Z_k$ at each detector. Source sky location fixes arrival time difference between detectors, but we have to marginalize over the geocentric arrival time. We evaluate this for every sample set of parameters including the geocentric arrival time (in practice limited to a small range of possible values) and the phase constant. For the overall amplitude, without informed theoretical priors we choose to simply maximize $\ln L$ with respect to $x_0$.

To compute the likelihood integral over the prior under the null hypothesis, we compute on an (RA, DEC) grid, perform a Montecarlo integration over $\iota$ and $\psi$, analytically marginalize over $x_0$ and $\phi_0$, and sum over all possible choices of arrival times. The Bayes score we compute is:
\begin{align}
\label{eq:Bext}
    B_{\rm ext}= \frac{\sum_{{\rm post}} \sum_{\rm ph} \sum_{{\rm time}} L|_{{\rm max}\ x_0}} { {N_{\rm post} N_{\rm ph}} \int \rmd\theta_{\rm ext} L},
\end{align}
where $N_{\rm post}$ is the number of posterior samples of GW170814 used and $N_{\rm ph}=4$ is the number of possible phase values given the phase of GW170814 and the geometric optic assumption. 

\begin{figure*}
     \centering
     \includegraphics[width=\linewidth]{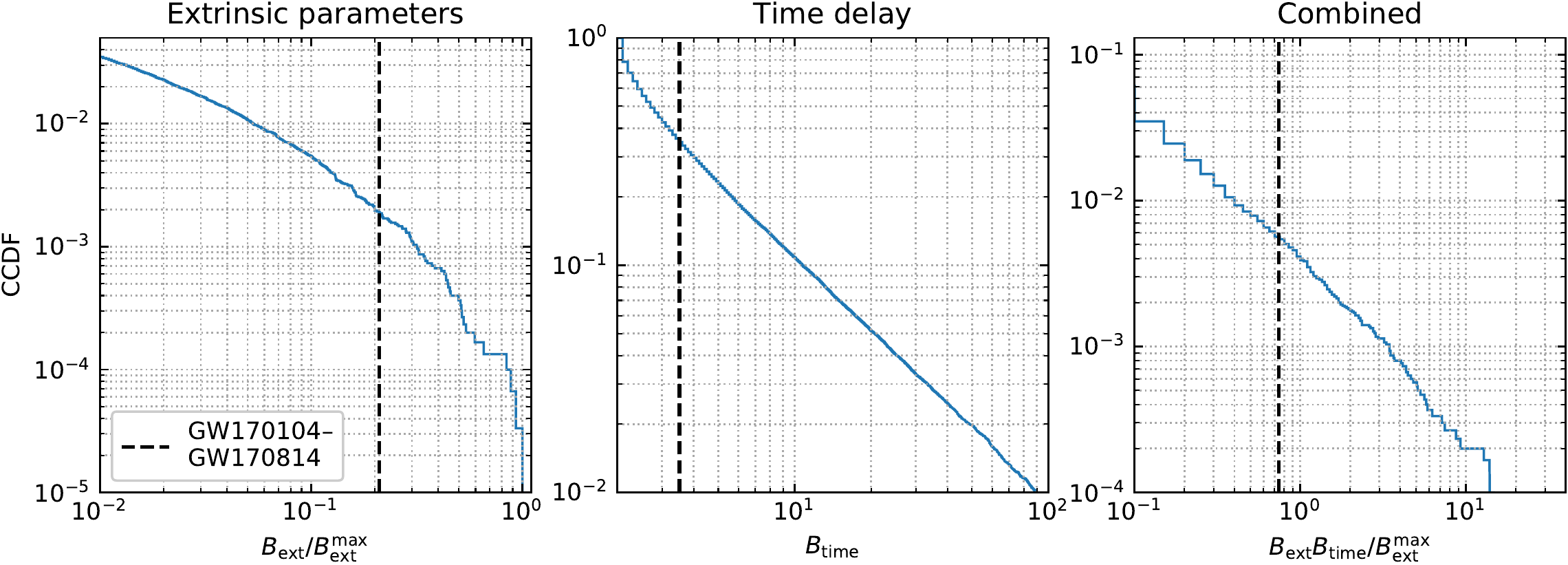}
     \caption{Significance for coincidence in extrinsic parameters and a long time delay between GW170814 and GW170104. Complementary cumulative distributions under the null hypothesis are shown for $B_{\rm ext}$ (left; Eq.~(\ref{eq:Bext})), $B_{\rm time}$ (center; Eq.~(\ref{eq:Btime})), and their product (right), with values observed for the pair GW170814 and GW170104 indicated by vertical dashed lines.
     }
     \label{fig:Bext}
 \end{figure*}

To compute a false alarm probability, we inject signals with extrinsic parameters drawn from the prior distribution. We set the amplitude so that on average injections have the same SNR as GW170104 has although the recovered SNRs expectedly fluctuate due to noise. \reffig{Bext} shows that only $2.1 \times 10^{-3}$ of the injections have a larger value for $B_{\rm ext}$ than what is measured for GW170104. If we disregard the Morse phase information, that is if both events are allowed to have independent phases, $1\times 10^{-2}$ of the injections have a larger $B_{\rm ext}$ than what is measured for GW170814. The Morse phase information therefore contributes about a factor of 5 reduction in the false alarm probability.\footnote{For a sanity check, we carry out a Monte Carlo test where only the phase of the samples of GW170814 and the sidereal hour of GW170104 are randomized, thus effectively shuffling only the RA. This procedure is simpler because it requires no injections into the data. In this case, $7\times 10^{-3}$ of the injections have a larger value of $B_{\rm ext}$ than what is measured for GW170104. There is roughly a factor of 6 reduction in the false alarm probability by accounting for the coincidence in DEC.}

\subsection{Time delay}

The time delay between GW170814 and GW170104 $\delta t \approx 222\,$d is rather long. We have argued that this could have been produced by group-scale or cluster-scale lenses and that their contribution to the strong lensing optical depth is not negligible. However, had we observed a much shorter time delay we would have been inclined to take that as additional evidence in favor of lensing~\cite{HarisGWLensingTechnique, HannukselaGWLensing}. To compute a Bayes factor for the time delay we would need a model for the expected delays from lensing. We could not do this without making many uncertain assumptions and choices, or use a realistic catalog of mock lenses~\cite{OguriGWLensing, RobertsonLensingPredictions}, so instead we will simply assume that time delays from galaxy to cluster lenses span a very large range and follow a log flat distribution. 

For unlensed events, we take the distribution of arrival time to be uniform throughout O2, neglecting detector sensitivity variation. The goal of this simplification is to estimate the order of magnitude of this effect. With the above simple choices, the Bayes factor for the time delay is simply:
\begin{align}
\label{eq:Btime}
    B_{\rm time} = \frac{T^2}{2\,\delta t\, (T-\delta t)},
\end{align}
where $\delta t$ is the time delay and $T$ is the entire duration of the observing run. To compute a false alarm probability, we sample from the prior distribution. As we show in \reffig{Bext}, there is only a \num{5.5e-3} chance that a combined value $B_{\rm ext}\,B_{\rm time}$ is larger than what is observed for GW170104 with respect to GW170814.

\subsection{Coherent score for finding sub-threshold images}
\label{ap:ThirdImageStatistic}

Associating GW170814 with GW170104 strongly constrains source and lens properties. Exploiting this information can substantially benefit the search for additional low-SNR images as the effective look-elsewhere effect can be reduced.
The most obvious piece of information to use are the intrinsic parameters, which simplifies the search to have a single template.
Less trivial, but just as important, is prediction of time delays and phase differences between detectors based on the source sky location and other extrinsic parameters inferred from GW170814 and GW170104, accounting for the Morse phase.

To find additional associated images, we first search all times in the LIGO O2 data with the following constraints:
\begin{enumerate}
    \item Squared SNR with the fiducial template (maximum likelihood template of GW170814) is larger than 47.
    \item Squared SNR with any other template does not exceed the squared SNR measured with the fiducial template by more than 5 units. This enforces compatible intrinsic parameters.
    \item Squared SNR at either of the LIGO detectors is larger than 16.
    \item Triggers in both LIGO detectors pass all our signal consistency checks.
\end{enumerate}
By applying exactly the same search procedure to unphysical time slides between detectors, we collect background candidates.
For all candidates and background candidates, we then apply the following coherent score to evaluate a final association score according to which we rank the triggers.

The coherent score is computed using:
\begin{equation}
    S \propto \sum_{s\in \Pi_{\rm pair}}{\sum_{\phi_M, t_0}{\int{\rmd D_L\, \mathcal{L}\left(s,\phi_M,t_0,D_L\right)\,P(D_L)}}},
\end{equation}
where the prior $P(D_L)$ is taken to be flat in the relevant SNR range. Experimenting with this prior, we checked that it does not affect candidate ranking, but only affects the overall normalization which is immaterial for the purpose of computing the FAP.

We generate background triggers worth of 2000 O2 observing runs. Only 26 of those have a score larger than what we find with GWC170620, corresponding to a FAP of $1.3 \times 10^{-2}$.

\section{Estimate of the lensing probability}
\label{ap:LensingExpectations}

For a crude estimate of the a priori probability of observing multiple lensed images, we follow the formalism of Ref.~\cite{Dai:2016igl}. We consider the differential detection rate of lensed BBH events as a function of observed chirp mass ($\tM$) and redshift ($\tz$), and express this as an integral over the flux magnification $\mu$:
\begin{equation}
    \frac{\rmd^3 N_{\rm sl} (\tM,\tz)}{\rmd\tM\, \rmd\tz\, \rmd t} = \int \rmd\ln \mu \frac{\rmd P_{\rm sl}(\mu,z)}{\rmd \ln \mu} \frac{\rmd^3 N (M,z)}{\rmd M\, \rmd z\, \rmd t} \left|\frac{\partial (M,z)}{\partial (\tM,\tz)}\right|,
\end{equation}
where ${\rmd P_{\rm sl}(\mu,z)}/{\rmd \ln \mu}$ gives the probability density of having a source at redshift $z$ strongly lensed with magnification $\mu$ and $|{\partial (M,z)}/{\partial (\tM,\tz)}|$ is the Jacobian of the variable transformation. The observed and true source redshifts are related by $D_L(\tz)= D_L(z) \mu^{-1/2}$ where $D_L$ is the luminosity distance. The differential event rate is related to the differential merger rate density (per unit comoving volume) as:
\begin{align}
    \frac{\rmd^3 N (M,z)}{\rmd M\, \rmd z\, \rmd t}
    &= \frac{\rmd^2 n (M,z)}{\rmd M\, \rmd t_s}
        \frac{1}{(1+z)} \frac{\rmd V}{\rmd z}, \\
    \frac{\rmd V}{\rmd z} &= \frac{4 \pi c\, \chi^2(z)}{H(z)},
\end{align}
with $H(z)$ the Hubble parameter and $\chi(z)=D_L(z)/(1+z)$ the comoving distance to redshift $z$. We will define the lensing probability as
\begin{equation}
    P_{\rm lensed} = \frac{{\rmd^3 N_{\rm sl} (\tM,\tz)}/{\rmd\tM \rmd\tz \rmd t}}{{\rmd^3 N (\tM,\tz)}/{\rmd\tM \rmd\tz \rmd t}}.
\end{equation}
Differentiating the numerator with respect to $\ln\mu$ while keeping the denominator, we can similarly define $\rmd P_{\rm lensed}/\rmd\ln\mu$, which satisfies $\int\,\rmd\ln\mu\,(\rmd P_{\rm lensed}/\rmd\ln\mu) = P_{\rm lensed}$.

For an order of magnitude estimate of this probability, we use the simple model of an singular isothermal sphere with an internal velocity dispersion $\sigma$. In this case we have: 
\begin{align}
    \frac{\rmd P_{\rm sl}(\mu,z_s)}{\rmd\ln \mu} &= P_{\rm sl}(z_s)\,\frac{8}{\mu^2},\\
    P_{\rm sl}(z_s) &= \int \rmd z_l\,\frac{\rmd^2 V}{\rmd z \rmd\Omega}
        \int \rmd\ln\sigma\,\frac{\rmd n}{\rmd\ln\sigma}\, \sigma_{\rm sl}(\sigma),\\
    \sigma_{\rm sl}(\sigma)
    &= \pi\,\theta_E^2=\pi \left[4 \pi \left(\frac{\sigma}{c}\right)^2 \frac{D_{LS}}{D_{S}}\right]^2,
\end{align}
where $D_{S}$ and $D_{LS}$ are the angular diameter distances to the source, and from the lens to the source, respectively. 
Toward low source redshift, $P_{\rm sl}(z_s)\propto z_s^3$. For our estimate, we use the approximate fitting formula reported in Ref.~\cite{OguriTransientLensing}:
\begin{align}
    P_{\rm sl}(z_s)\approx A \ \frac{z_s^3}{(1 + 0.41\, z_s^{1.1})^{2.7}}.
\end{align}
The contribution from galaxy-scale lenses, as reported in Ref.~\cite{OguriTransientLensing}, translates into $A = \num{5e-4}$. To estimate the contribution of group-/cluster-scale lenses, we consider the mass function of dark matter halos. We use the mass function from \cite{Tinker:2008ff} for the Planck cosmology and use the scaling relation between mass and internal velocity dispersion reported in Ref.~\cite{Evrard:2007py}. These inputs lead to
$A=\num{6e-4}$ when the integral $P_{\rm sl}(z_s)$ is evaluated with the halo mass function at zero redshift, appropriate for the low redshift lenses relevant for our candidate event.

\begin{figure}
     \centering
     \includegraphics[width=\linewidth]{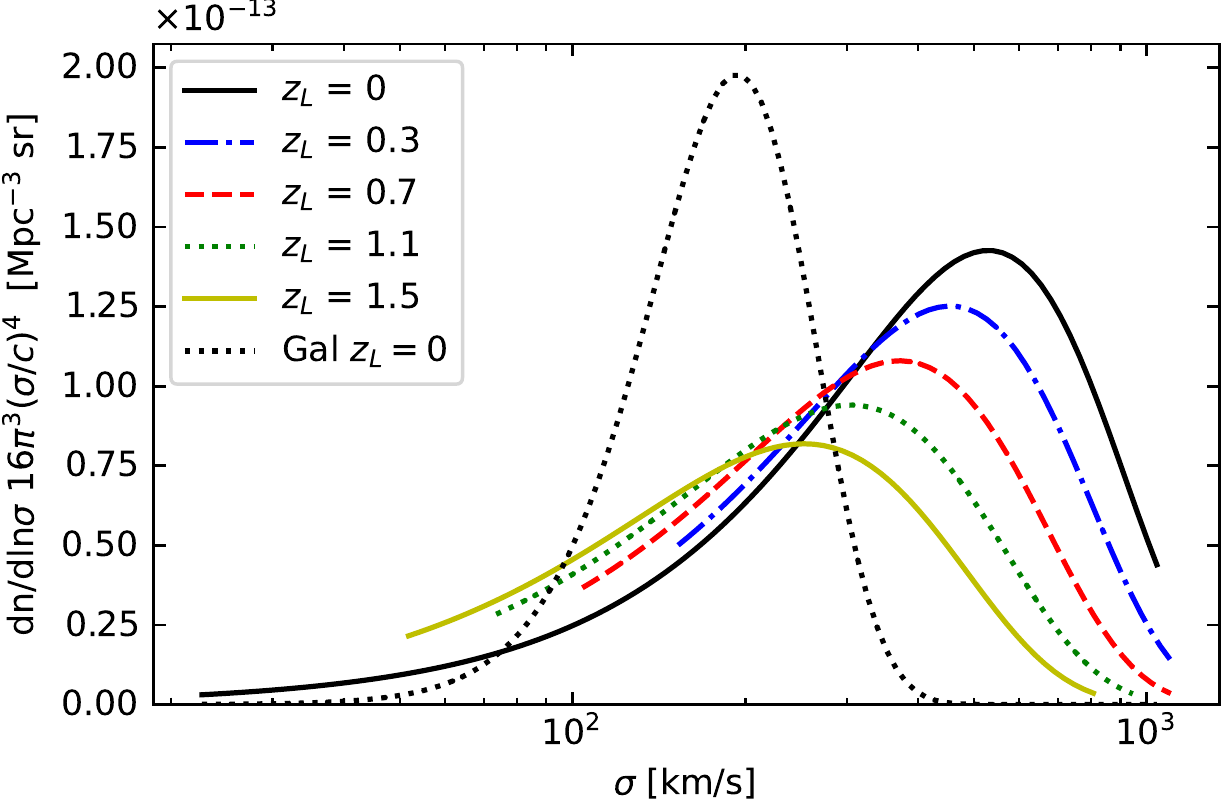}
     \caption{Differential contribution to the strong lensing cross section with respect to the lens internal velocity dispersion. The dotted black line is for galaxy-scale lenses in the present-day Universe, while the other curves are for dark-matter halos, for a number of lens redshifts.
     In particular for low source redshifts, an important contribution to the lensing cross-section is from galaxy clusters. We expect that a non-negligible fraction of strongly lensed events occurring during O2 would be caused by complex lenses of galaxy cluster scale and would have long time delays between lensed images.}
     \label{fig:velocity-dispersion}
 \end{figure}

In Fig.~\ref{fig:velocity-dispersion}, we plot the integrand $\rmd n/\rmd\ln\sigma$. The integrand peaks around $\sigma=\SI{600}{\kilo\meter/\second}$, with significant support all the way to $\sigma=\SI{1000}{\kilo\meter/\second}$. We note that the SIS model overestimates the inner total matter content in galaxy clusters. Studies of massive and relaxed galaxy clusters~\cite{del2014flat} suggest that the inner slope of the total density profile is often closer to that of the Navarro-Frenk-White (NFW) profile~\cite{navarro1997universal}. Hence the actual lensing cross section from galaxy clusters is likely to be less than indicated in Fig.~\ref{fig:velocity-dispersion}. 

Still, the contribution from lenses of high velocity dispersion is significant. Indeed, cases of multiple images of lensed quasar with long time delays are not rare~\cite{rathna2015h0}. In particular, for halos at low redshifts the integral receives sizable contribution from lenses whose internal velocity dispersions are significantly higher than those of galaxies. For these type of lenses, long time delays are natural,
\begin{multline} 
    \Delta t_{\rm SIS} = \SI{203}{days} \times{}  \\
     \frac{D_{LS}\,D_{L}/D_{S}}{\SI{230}{Mpc}}
     \left(\frac{\sigma}{\SI{650}{\kilo\meter/\second} }\right)^4
     \left(\frac{\mu}{10}\right)^{-1} \frac{1+z_l}{1.07}.
\end{multline}

Because the results in Ref.~\cite{OguriTransientLensing} are only based on galaxy-scale lenses with lower internal velocity dispersions than what is reported here, the derived time delays are shorter. Our estimate indicates that for low redshift sources, the contribution to strong lensing optical depth from group-/cluster-scale lenses is non-negligible and thus a $\sim 200$ day time delay is not at all out of the question. 

To carry out the calculation, we need the mass and time dependence of ${\rmd^2 n (M,z)}/{\rmd M \rmd t_s}$, which are currently not well constrained by gravitational wave data. For simplicity, we adopt the parametric form
\begin{align}
    \frac{\rmd^2 n (M,z)}{\rmd M \rmd t_s}\propto M^{-\alpha_M}\, (1+z)^{\alpha_z} \propto \left(\frac{1+z}{1+\tz}\right)^{\alpha_M+\alpha_z}.
\end{align}
The power-law indices $\alpha_M$ and $\alpha_z$ are constrained in the analysis of Ref.~\cite{LIGOScientific:2018jsj}. A value $\alpha_M \sim 2$ is reported. Although a preference for positive $\alpha_z$ is noted, uncertainty for the value of this parameter is large: $\alpha_z\approx 6.5 \pm 9$. For simplicity, we take $\alpha_M + \alpha_z = 4$. Given the preferred low source redshift of our candidate, this choice is not crucial in determining the order of magnitude of the strong lensing rate unless a very rapid redshift evolution is assumed. 

Let us consider GW170814 with $\tz_{170814}\approx 0.11$ and assume a magnification factor $\mu\sim 10$. In this case,
\begin{eqnarray}
     z_s &\approx 0.3,  \\
     \frac{\rmd P_{\rm sl}(\mu,z)}{\rmd \ln \mu} &\approx 10^{-6}, \\
     \frac{{\rmd V}/{\rmd z}(z_s)} {{\rmd V}/{\rmd z}(\tz)} &\approx 6, \\
     \left|\frac{\partial (M,z)}{\partial (\tM,\tz)}\right| &\approx 2.2, \\
     \left(\frac{1+z_s}{1+\tz}\right)^4 &\approx 1.4.
\end{eqnarray}
giving a combined
\begin{align}
    \frac{\rmd P_{\rm lensed}}{\rmd\ln \mu} \approx \num{2e-5}.
\end{align}
Integrated over all magnifications, we get $P_{\rm lensed}(\tz= 0.11)\approx \num{2e-4}$. Restricted to $\mu < 10$, we only get $P_{\rm lensed}(\tz= 0.11,\,\mu < 10)\approx \num{2e-5}$. In this simple model and for the low observed source redshift assumed, lensing at high magnification makes a dominant contribution. 

It is important to note that $P_{\rm lensed}$ grows strongly with an increasing apparent source redshift. For a BBH source with a higher chirp mass typically observed from a higher redshift, say $\tz= 0.5$, $P_{\rm lensed}(\tz= 0.5)\approx \num{2e-3}$.

\bibliographystyle{apsrev4-1-etal}
\bibliography{gw}

\end{document}